\documentclass[fleqn]{mat01}
\usepackage{times,mathtimy,amssymb,boxit}
\begin{document}

\setcounter{page}{207}
\firstpage{207}

\font\xx=msam5 at 9pt
\def\ab{\mbox{\xx{\char'03}}}

\font\sa=tibi at 10.4pt
\def\d{\hbox{d}}

\font\nars=mtmib at 13.5pt
\def\bbet{\mbox{\nars \char'014}}

\def\defi{\trivlist\item[\hskip\labelsep{\bf DEFINITION.}]}
\def\remark{\trivlist\item[\hskip\labelsep{\it Remark.}]}
\def\remarks{\trivlist\item[\hskip\labelsep{\it Remarks}]}
\def\noot{\trivlist\item[\hskip\labelsep{{\it Note.}}]}
\def\thoe{\trivlist\item[\hskip\labelsep{{\bf Theorem}}]}

\newtheorem{dets}{\rm DEFINITION}
\newtheorem{theo}{\bf Theorem}

\title{Bounds on the phase velocity in the linear instability of viscous
shear flow problem in the $\beta$-plane}

\markboth{R~G~Shandil and Jagjit Singh}{Viscous linear shear flow instability}

\author{R~G~SHANDIL and JAGJIT SINGH$^{*}$}

\address{Department of Mathematics, H.P. University, Shimla 171 005,
India\\
\noindent $^{*}$Sidharth Govt. Degree College, Nadaun, Distt Hamirpur 177 033,
India
}

\volume{113}

\mon{May}

\parts{2}

\Date{MS received 4 July 2002; revised 23 November 2002}

\begin{abstract}
Results obtained by Joseph ({\it J. Fluid Mech.} {\bf 33} (1968) 617) for
the viscous parallel shear flow problem are extended to the problem of
viscous parallel, shear flow problem in the beta plane and a sufficient
condition for stability has also been derived.
\end{abstract}

\maketitle

\section{Introduction}

Parallel shear flows problem is a classical hydrodynamic instability
problem and continues to attract attention of researchers
\cite{BSG,BSC,DH,FH}. Kuo \cite{KHL} considered slightly general case of a homogeneous, inviscid,
parallel shear flow problem in the $\beta$-plane. He obtained the
linearized perturbation equation for this problem and derived an
extension of the celebrated Rayleigh's \cite{RS} inflexion point criterion for
inviscid, homogeneous parallel shear flows. Fjortoft's \cite{FR} instability
criterion for inviscid, homogeneous parallel shear flows has also been
extended for such barotropic flows (see \cite{HFJ}). Hickernel reviewed various
results obtained for this problem and obtained new upper bound on the
growth rate of the temporally growing modes of the linearized equations
of motion for this problem. Joseph \cite{JDD} obtained bounds for the
eigenvalues of the Orr--Sommerfeld's equation. Derivation of bounds is
of importance because exact solution of the problem is not obtainable in
closed form. In the present paper, we extend the results of Joseph
\cite{JDD}
to the problem of viscous parallel shear flow problem in the $\beta$-plane and also obtain a sufficient condition for stability.

The governing equation for Kuo's \cite{KHL} problem for the linear stability of
an incompressible, viscous, parallel shear flow in the $\beta$-plane is
given by (cf. \cite{HFJ})
\begin{equation}\label{eq1}
\big(U - c\big)\big(D^{2} - \alpha^{2}\big) \phi - \big( D^{2} U -
\beta\big)\phi = \frac{\big(D^{2} - \alpha^{2}\big)^{2}}{i\alpha R}\phi
\end{equation}
and the associated boundary conditions are that $\phi$ must vanish on
the rigid walls which may recede to $\pm \infty$ in the limiting cases
and thus
\begin{equation}\label{eq2}
\phi = D \phi = 0 \quad \hbox{at} \quad z = - 1 \quad \hbox{and} \quad z
= 1,
\end{equation}
where $z$ is the real independent variable such that $-1\leq z \leq 1, D
\equiv \d/\d z, \alpha$ is a real constant and denotes the wave number,
$c = c_{r} + ic_{i}$ is the complex wave velocity, $U(z)$ is a twice
continuously differentiable function of $z$ and denotes the prescribed
basic velocity distribution while the dependent variable $\phi(z)$ is,
in general, a complex valued function of $z$ and denotes the $z$
component of velocity distribution of the parallel flow, the parameter
$\beta$ is the derivative of the Coriolis force in latitudinal direction
\cite{HFJ} and $R$ is the Reynolds number. Here, we note that eq.~\eqref{eq1} for
$\beta = 0$ reduces to the celebrated Orr--Sommerfeld's equation.

\section{Mathematical analysis}

We prove the following theorems:

\begin{theo}[\!]
If $(\phi, c, \alpha^{2})$ is a solution of eigenvalue problem
prescribed by eqs~\eqref{eq1} and \eqref{eq2} for given values of $R$ and $\beta$, then
\begin{equation}\label{eq3}
I_{2}^{2} + 2\alpha^{2} I_{1}^{2} + \alpha^{4} I_{0}^{2} = i \alpha c R
\big(I_{1}^{2} + \alpha^{2} I_{0}^{2}\big) - i\alpha RQ,
\end{equation}
where we take
\begin{equation}\label{eq4}
I_{n}^{2} = \int_{-1}^{1} |D^{n} \phi|^{2} \,\hbox{\rm \d} z \qquad
(n = 0, 1, 2)
\end{equation}
and
\begin{align}
Q &= \int_{-1}^{1} \lbrace U(\vert
D\phi\vert^{2} + \alpha^{2}
\vert\phi\vert^{2}) + (D^{2} U -
\beta)\vert\phi\vert^{2} \rbrace\,\hbox{\rm \d} z\nonumber\\
&\quad \ +
\int_{-1}^{1} DU \big(D\phi\big)\phi^{*} \,\hbox{\rm \d} z.\label{eq5}
\end{align}
\end{theo}

\begin{proof}
Multiplying eq.~\eqref{eq1} by $\phi^{*}$ (the complex conjugate of $\phi$)
throughout and integrating the resulting equation over the vertical
range of $z$, using the boundary conditions (2), we derive
\begin{equation}\label{eq6}
I_{2}^{2} + 2\alpha^{2} I_{1}^{2} + \alpha^{4} I_{0}^{2} = i \alpha cR
\left(I_{1}^{2} + \alpha^{2} I_{0}^{2}\right) - i \alpha R Q.
\end{equation}
This proves the theorem.
\end{proof}

\begin{theo}[\!]
If $(\phi, c, \alpha^{2})$ is a solution of eigenvalue problem
prescribed by eqs~{\rm \eqref{eq1}} and {\rm \eqref{eq2}} for given values of $R$ and $\beta$, then
\begin{equation}\label{eq7}
c_{r} = \frac{\hbox{\rm Re}(Q)}{I_{1}^{2} + \alpha^{2} I_{0}^{2}}
\end{equation}
and
\begin{equation}\label{eq8}
c_{i} = \frac{1}{\left(I_{1}^{2} + \alpha^{2} I_{0}^{2}\right)}
\left\lbrace \hbox{\rm Im} (Q) - \frac{\left(I_{2}^{2} + 2\alpha^{2}
I_{1}^{2} + \alpha^{4}I_{0}^{2}\right)}{\alpha R}\right\rbrace,
\end{equation}
where
\begin{equation}\label{eq9}
\hbox{\rm Re} (Q) = \int_{-1}^{1} \left\lbrace U(\vert
D\phi\vert^{2} + \alpha^{2}
\vert\phi\vert^{2}) + \left(\frac{D^{2}U}{2} -
\beta\right) \vert\phi\vert^{2} \right\rbrace \,\hbox{\rm \d}z
\end{equation}
and
\begin{equation}\label{eq10}
\hbox{\rm Im} (Q) = \frac{i}{2} \int_{-1}^{1} \left\lbrace DU
\left\lbrace\phi \left(D\phi^{*}\right) -
\big(D\phi\big)\phi^{*}\right\rbrace\right. \,\hbox{\rm \d} z.
\end{equation}
\end{theo}

\begin{proof}
Equating the real part of both sides of eq.~\eqref{eq3}, we obtain
\begin{equation*}
I_{2}^{2} + 2\alpha^{2} I_{1}^{2} + \alpha^{4} I_{0}^{2} = - \alpha
c_{i} R \left( I_{1}^{2} + \alpha^{2} I_{0}^{2}\right) +\alpha
R\ \hbox{\rm Im}(Q)
\end{equation*}
or
\begin{equation}\label{eq11}
c_{i} = \frac{1}{\big(I_{1}^{2} + \alpha^{2} I_{0}^{2}\big)}
\left\lbrace \hbox{Im} (Q) - \frac{\left(I_{2}^{2} + 2\alpha^{2}
I_{1}^{2} + \alpha^{4} I_{0}^{2}\right)}{\alpha R}\right\rbrace.
\end{equation}
Now, equating the imaginary parts of both sides of eq.~\eqref{eq3}, we obtain
\begin{equation}\label{eq12}
c_{r} = \frac{\hbox{Re}(Q)}{\big(I_{1}^{2} + \alpha^{2} I_{0}^{2}\big)}.
\end{equation}
This proves the theorem.
\end{proof}

\begin{theo}[\!]
If $(\phi, c, \alpha^{2})$ is a solution of eigenvalue problem
prescribed by eqs~{\rm \eqref{eq1}} and {\rm \eqref{eq2}} for given values of $R$ and $\beta${\rm ,} then
\begin{equation}\label{eq13}
c_{i} \leq \frac{q}{2\alpha} - \frac{1}{\alpha R} \left(
\frac{\pi^{2}}{4} + \alpha^{2}\right),
\end{equation}
where $q = \max \lvert DU\rvert$ on $[-1,1]$.
\end{theo}

\begin{proof}
Taking modulus of both sides of eq.~\eqref{eq10}, we obtain
\begin{align}
\hbox{Im}\,(Q) \leq \big\vert\hbox{Im}\,(Q)\big\vert &= \left\vert
\frac{i}{2} \int_{-1}^{1} DU\left\lbrace \phi
\big(D\phi^{*}\big) - \big(D \phi\big) \phi^{*}\right\rbrace \,\d
z\right\vert\nonumber\\
&\leq \int_{-1}^{1} \vert DU\Vert\phi\Vert
D\phi \vert \,\d z \leq q \int_{-1}^{1} \vert\phi
\Vert D\phi \vert \,\d z\nonumber\\
&\leq q \sqrt{\int_{-1}^{1} \vert
D\phi\vert^{2}\,\d z} \quad \sqrt{\int_{-1}^{1}
\vert\phi\vert^{2} \,\d z} = qI_{1}I_{0}.\label{eq14}\\
&\qquad \quad \hbox{(On using Schwartz's inequality)}\nonumber
\end{align}
On using inequality \eqref{eq14} in eq.~\eqref{eq11}, we get
\begin{equation}\label{eq15}
c_{i} \leq \frac{1}{\big(I_{1}^{2} + \alpha^{2} I_{0}^{2}\big)}
\left\lbrace qI_{1} I_{0} - \frac{\big(I_{2}^{2} + 2\alpha^{2} I_{1}^{2}
+ \alpha^{4} I_{0}^{2}\big)}{\alpha R} \right\rbrace.
\end{equation}
Clearly
\begin{equation}\label{eq16}
I_{1}^{2} + \alpha^{2} I_{0}^{2} \geq 2 \alpha I_{1} I_{0}.
\end{equation}
Using the isoperimetric inequalities
$\displaystyle I_{2}^{2} \geq (\pi^{2}/{4}) I_{1}^{2}$ and $\displaystyle
I_{1}^{2} \geq (\pi^{2}/{4}) I_{0}^{2}$, we have
\begin{align}
I_{2}^{2} + 2\alpha^{2} I_{1}^{2} + \alpha^{4} I_{0}^{2} &=
\big(I_{2}^{2} + \alpha^{2} I_{1}^{2}\big) + \alpha^{2} \big(I_{1}^{2} +
\alpha^{2} I_{0}^{2}\big)\nonumber\\[.5pc]
&\geq \left( \frac{\pi^{4}}{4} + \alpha^{2}\right)
I_{1}^{2} + \alpha^{2} \left( \frac{\pi^{4}}{4} +
\alpha^{2}\right) I_{0}^{2}\nonumber\\[.5pc]
&= \left( \frac{\pi^{4}}{4} + \alpha^{2}\right)
\big(I_{1}^{2} + \alpha^{2} I_{0}^{2}\big).\label{eq17}
\end{align}
Now, from inequalities \eqref{eq15}, \eqref{eq16} and \eqref{eq17}, we get
\begin{equation}\label{eq18}
c_{i} \leq \frac{q}{2\alpha} - \frac{1}{\alpha R} \left(
\frac{\pi^{2}}{4} + \alpha^{2}\right).
\end{equation}
This proves theorem~3.

Inequality \eqref{eq18} gives an upper bound on the growth (or decay) rate of a
disturbance of wave number $\alpha$.
\end{proof}

\begin{theo}[\!]
If $(\phi, c, \alpha^{2})$ is a solution of eigenvalue problem
prescribed by eqs~{\rm \eqref{eq1}} and {\rm \eqref{eq2}} for given values of $R$ and $\beta${\rm ,} and
if $\alpha R$ is small enough then the flow will be stable.
\end{theo}

\begin{proof}
Inequality \eqref{eq15} can be written as
\begin{equation*}
\big(I_{1}^{2} + \alpha^{2} I_{0}^{2}\big) \alpha Rc_{i} \leq
\big\lbrace \alpha Rq I_{1} I_{0} - \big(I_{2}^{2} + 2\alpha^{2}
I_{1}^{2} + \alpha^{4}I_{0}^{2}\big)\big\rbrace.
\end{equation*}
Thus if $\alpha R$ is small enough then $c_{i}$ will be negative and the
flow will be stable.
\end{proof}

\begin{theo}[\!]
If $(\phi, c, \alpha^{2})$ is  a solution of eigenvalue problem
prescribed by  eqs~{\rm \eqref{eq1}} and {\rm \eqref{eq2}} for given values of $R$ and $\beta$ then
the following inequalities hold good:
\end{theo}

{\it
\begin{enumerate}
\leftskip .4pc
\renewcommand{\labelenumi}{\rm (\roman{enumi})}
\item If $D^{2} U_{\min}/{2} \geq \beta,$  then
\begin{equation}\label{eq19}
U_{\min} < c_{r} < U_{\max} + \frac{4\displaystyle \left(
\frac{D^{2}U_{\max}}{2} - \beta\right)}{\pi^{2} + 4\alpha^{2}}.
\end{equation}

\item  If $D^{2}U_{\min}/2 \leq \beta \leq D^{2} U_{\max}/2,$ then
\begin{equation}\label{eq20}
U_{\min} + \frac{4\displaystyle \left( \frac{D^{2}U_{\min}}{2} -
\beta\right)}{\pi^{2} + 4\alpha^{2}} < c_{r} < U_{\max} + \frac{4\displaystyle \left( \frac{D^{2}U_{\max}}{2} -
\beta\right)}{\pi^{2} + 4\alpha^{2}}.
\end{equation}

\item If $D^{2} U_{\max}/{2} \leq \beta,$ then
\begin{equation}\label{eq21}
U_{\min} + \frac{4\displaystyle \left( \frac{D^{2}U_{\min}}{2} -
\beta\right)}{\pi^{2} + 4\alpha^{2}} < c_{r} < U_{\max},
\end{equation}
\end{enumerate}
where $f_{\max}$ and $f_{\min}$ are respectively the $\max f(z)$ and
$\min f(z)$ on $z \in [-1,1]$.
}

\begin{proof}
From eq.~\eqref{eq7} we have
\begin{equation}\label{eq22}
U_{\min} + \frac{\left(\displaystyle \frac{D^{2}U_{\min}}{2} -
\beta\right) I_{0}^{2}}{\big(I_{1}^{2} + \alpha^{2}
I_{0}^{2}\big)} < c_{r} < U_{\max} + \frac{\left(\displaystyle \frac{D^{2}U_{\max}}{2} -
\beta\right) I_{0}^{2}}{\big(I_{1}^{2} + \alpha^{2}
I_{0}^{2}\big)}.
\end{equation}
Now, we prove part (i) of the theorem. Since, $D^{2} U_{\min}/{2} \geq
\beta$, therefore
\begin{equation}\label{eq23}
U_{\min} < U_{\min} + \frac{\left(\displaystyle \frac{D^{2}U_{\min}}{2} -
\beta\right) I_{0}^{2}}{\big(I_{1}^{2} + \alpha^{2}
I_{0}^{2}\big)} < c_{r}.
\end{equation}
By Rayleigh--Ritz inequality, we have
\begin{equation}\label{eq24}
I_{1}^{2} \geq \frac{\pi^{2}}{4} I_{0}^{2} \quad (\hbox{since} \ \phi(-
1) = \phi(1) = 0).
\end{equation}
Using inequality \eqref{eq24}, we obtain
\begin{equation}\label{eq25}
\frac{I_{0}^{2}}{I_{1}^{2} + \alpha^{2} I_{0}^{2}} \leq \frac{4}{\pi^{2}
+ 4 \alpha^{2}}.
\end{equation}
Therefore
\begin{equation}\label{eq26}
\frac{\left(\displaystyle \frac{D^{2}U_{\max}}{2} -
\beta\right) I_{0}^{2}}{I_{1}^{2} + \alpha^{2}
I_{0}^{2}} \leq \frac{\displaystyle 4\left( \frac{D^{2} U_{\max}}{2} -
\beta\right)}{\pi^{2} + 4\alpha^{2}}
\end{equation}
since, $\beta \leq D^{2} U_{\min}/{2} \leq D^{2} U_{\max}/{2}$.
Therefore
\begin{equation}\label{eq27}
c_{r} < U_{\max} + \frac{\displaystyle \left( \frac{D^{2} U_{\max}}{2} -
\beta\right) I_{0}^{2}}{I_{1}^{2} + \alpha^{2} I_{0}^{2}} \leq
U_{\max} + \frac{4 \displaystyle \left( \frac{D^{2} U_{\max}}{2} -
\beta\right)}{\pi^{2} + 4\alpha^{2}}.
\end{equation}
On using \eqref{eq23} and \eqref{eq27}, we obtain
\begin{equation}\label{eq28}
U_{\min} <c_{r}< U_{\max} + \frac{4\displaystyle \left(
\frac{D^{2} U_{\max}}{2} -\beta\right)}{\pi^{2} + 4\alpha^{2}}.
\end{equation}
This completes the proof of part (i).

In case (ii), we have $D^{2} U_{\min}/2 \leq \beta\leq D^{2}
U_{\max}/2$, thus $(D^{2} U_{\min}/{2}) - \beta \leq 0$. Therefore, from
the inequality \eqref{eq25} we have
\begin{equation}\label{eq29}
\frac{\left( \displaystyle \frac{D^{2} U_{\min}}{2} -
\beta\right) I_{0}^{2}}{I_{1}^{2} + \alpha^{2} I_{0}^{2}} \geq
\frac{4 \left( \displaystyle \frac{D^{2} U_{\min}}{2} -
\beta\right)}{\pi^{2} + 4\alpha^{2}}
\end{equation}
or
\begin{equation}\label{eq30}
U_{\min} + \frac{\left( \displaystyle \frac{D^{2}U_{\min}}{2} -
\beta\right) I_{0}^{2}}{I_{1}^{2} + \alpha^{2} I_{0}^{2}} \geq
U_{\min} + \frac{4\left( \displaystyle \frac{D^{2} U_{\min}}{2} -
\beta\right)}{\pi^{2} + 4\alpha^{2}}.
\end{equation}
Now, using inequalities \eqref{eq22} and \eqref{eq30} we get
\begin{equation}\label{eq31}
U_{\min} + \frac{4 \displaystyle \left( \frac{D^{2} U_{\min}}{2} -
\beta\right)}{\pi^{2} + 4\alpha^{2}} < c_{r}.
\end{equation}
On multiplying inequality \eqref{eq25} throughout by $((D^{2}
U_{\max}/{2}) - \beta) (\geq 0)$, adding $U_{\max}$ to
the resulting inequality throughout and using inequalities \eqref{eq22} and \eqref{eq31}
we get
\begin{equation}\label{eq32}
U_{\min} + \frac{4 \displaystyle \left( \frac{D^{2} U_{\min}}{2} -
\beta\right)}{\pi^{2} + 4\alpha^{2}} < c_{r} < U_{\max} + \frac{4
\displaystyle \left( \frac{D^{2}U_{\max}}{2} - \beta\right)}{\pi^{2} +
4\alpha^{2}}.
\end{equation}
This proves part (ii) of the theorem.

In case (iii) we have $((D^{2} U_{\max}/2\big) - \beta) \leq 0$, therefore
\begin{equation}\label{eq33}
U_{\max} + \frac{\displaystyle \left( \frac{D^{2} U_{\max}}{2} -
\beta\right) I_{0}^{2}}{I_{1}^{2} + \alpha^{2} I_{0}^{2}} \leq
U_{\max}.
\end{equation}
On multiplying inequality \eqref{eq25} throughout by $((D^{2}
U_{\min}/{2}\big) - \beta) (\leq 0)$, adding $U_{\min}$
throughout to the resulting inequality and using inequalities \eqref{eq22} and
\eqref{eq33} we obtain
\begin{equation}\label{eq34}
U_{\min} + \frac{4\left( \displaystyle \frac{D^{2}U_{\min}}{2} -
\beta\right)}{\pi^{2} + 4 \alpha^{2}} < c_{r} < U_{\max}.
\end{equation}
This establishes part (iii) of the theorem and this completes the proof
of Theorem~5. It is to be noted here that for $\beta = 0$ in eq.~\eqref{eq1}, we
get the eigenvalue problem considered by Joseph \cite{JDD} and in this case
results of Theorem~5 reduce to the results obtained by Joseph.
\end{proof}

\section*{Acknowledgement}

We are grateful to the learned referee for his valuable comments, which
resulted in the improvement of this paper.

\end{document}